\documentclass[prl, twocolumn,superscriptaddress,nobibnotes]{revtex4}
\usepackage{graphicx}
\usepackage{amssymb}
\usepackage{amsmath}
\usepackage{bm}
\usepackage{color}
\usepackage{float}
\usepackage{setspace}
\usepackage{physics}
\usepackage{subfigure}
\usepackage{lipsum}

\begin{document}
	
	\title{Electrically controlling vortices in a neutral exciton \\ polariton condensate at room temperature}

	\author{Xiaokun Zhai}
	\affiliation{Tianjin Key Laboratory of Molecular Optoelectronic Science, Department of Physics, School of Science, Tianjin University, Tianjin 300072, China} 
	\affiliation{Institute of Molecular Plus, Tianjin University, Tianjin 300072, China}
	
	\author{Xuekai Ma}
	
	\affiliation{Department of Physics and Center for Optoelectronics and Photonics Paderborn (CeOPP), Universit\"{a}t Paderborn, Warburger Strasse 100, 33098 Paderborn, Germany}
	
	\author{Ying Gao}
	\affiliation{Tianjin Key Laboratory of Molecular Optoelectronic Science, Department of Physics, School of Science, Tianjin University, Tianjin 300072, China} 
	\affiliation{Institute of Molecular Plus, Tianjin University, Tianjin 300072, China}
	
	\author{Chunzi Xing}
	\affiliation{Tianjin Key Laboratory of Low Dimensional Materials Physics and Preparing Technology, School of Science, Tianjin University, Tianjin 300072, China}
	
	\author{Meini Gao}
	\affiliation{Tianjin Key Laboratory of Low Dimensional Materials Physics and Preparing Technology, School of Science, Tianjin University, Tianjin 300072, China}
	
	\author{Haitao Dai}
	\affiliation{Tianjin Key Laboratory of Low Dimensional Materials Physics and Preparing Technology, School of Science, Tianjin University, Tianjin 300072, China}
	
	\author{Xiao Wang}
	\affiliation{College of Materials Science and Engineering, Hunan University, Changsha 410082, China}
	
	\author{Anlian Pan}
	\affiliation{College of Materials Science and Engineering, Hunan University, Changsha 410082, China}
	
	\author{Stefan Schumacher}
	\affiliation{Department of Physics and Center for Optoelectronics and Photonics Paderborn (CeOPP), Universit\"{a}t Paderborn, Warburger Strasse 100, 33098 Paderborn, Germany}
	\affiliation{Wyant College of Optical Sciences, University of Arizona, Tucson, AZ 85721, USA}
	
	\author{Tingge Gao}
	
	\affiliation{Tianjin Key Laboratory of Molecular Optoelectronic Science, Department of Physics, School of Science, Tianjin University, Tianjin 300072, China} 
	\affiliation{Institute of Molecular Plus, Tianjin University, Tianjin 300072, China}
	
	\begin{abstract}
		{Manipulating bosonic condensates with electric fields is very challenging as the electric fields do not directly interact with the neutral particles of the condensate. Here we demonstrate a simple electric method to tune the vorticity of exciton polariton condensates in a strong coupling liquid crystal (LC) microcavity with CsPbBr$_3$ microplates as active material at room temperature. In such a microcavity, the LC molecular director can be electrically modulated giving control over the polariton condensation in different modes. For isotropic non-resonant optical pumping we demonstrate the spontaneous formation of vortices with topological charges of +1, +2, -2, and -1. The topological vortex charge is controlled by a voltage in the range of 1 to 10 V applied to the microcavity sample. This control is achieved by the interplay of a built-in potential gradient, the anisotropy of the optically active perovskite microplates, and the electrically controllable LC molecular director in our system with intentionally broken rotational symmetry. Besides the fundamental interest in the achieved electric polariton vortex control at room temperature, our work paves the way to micron-sized emitters with electric control over the emitted light's phase profile and quantized orbital angular momentum for information processing and integration into photonic circuits.}
	\end{abstract}
	
	\maketitle
	
	Vortices are topological defects with phase winding of 2N$\pi$ around an intensity or density minimum, the so-called vortex core (N is an integer and characterizes the topological charge of a vortex). Such vortices have attracted considerable attention in many different areas of physics and beyond, for example in cold atom condensates \cite{1,1-2,1-3}, in superconductors \cite{2,2-2}, in nonlinear optics \cite{KivsharBook}, and in hybrid light-matter systems, e.g., exciton polaritons \cite{lagoudakis1, lagoudakis2}. The potential to tune the topological charges of vortices attracts much attention as it can be used in optical communication, information processing, optical signal sensing and trapping, and sources of topological light. In most systems, however, the vortex charge is difficult to be controlled or modified. Recently, it was found that optical vortex charge control can be realized by engineering the non-hermiticity in a photonic system \cite{tunable vortex laser}. Swapping the topological charge of a photonic vortex can also be achieved by optically spin polarizing the gain material \cite{Amo vortex}. However, direct electrical control of the topological vortex charge, which would be vital for integration into optoelectronic chips with vortices as information carriers or for integrated emitters with adjustable orbital angular momentum of the emitted light, is still elusive. 
	
	Compared with cold atom condensates and purely photonic nonlinear systems, exciton polaritons appear as a promising platform to electrically control the vorticity of the emitted light. In the latter systems, an electric current or voltage can be directly applied in a specially designed microcavity with doped DBR or conductive ITO layers. Exciton polaritons arise from the strong coupling of excitons and cavity photon modes. Being composite bosons, a bosonic condensation process can be observed when they are pumped non-resonantly by external optical fields or electric currents \cite{polariton BEC1, polariton BEC2, polariton electrical laser}. Vortices in exciton polariton condensates can be pinned to the disorder or defects in semiconductor microcavities \cite{lagoudakis1, lagoudakis2}. In addition, judiciously designed potential traps can select specifically charged vortex modes near exceptional points \cite{EP vortex}. Confined vortices can be observed in coupled micropillar structures \cite{Amo PRX}. Optically switching the topological charge of  polariton vortices was reported by utilizing a second off-resonant control beam \cite{Ma xuekai}. It has been demonstrated that electric fields influence polariton nonlinearities \cite{Tsintzos}, however, the direct electrical manipulation of polariton vortices has not been reported. 
	
	In microcavities, it was shown that the anisotropy of the cavity layer may act as an effective gauge field \cite{anisotropy GaAs} and significantly influences the polariton condensate distribution. In this context it is interesting to ask whether it is possible to electrically control the anisotropy of the microcavity and further influence the distribution of the polariton condensate. Utilizing a molecular LC inserted into the microcavity, the refractive index of the cavity layer can be easily tuned by electrically controlling the LC molecular director. As a consequence, the anisotropic photonic or polariton mode can be manipulated \cite{1-liquid crystal_science, Yao RD}.

	In this work, we fabricate a microcavity filled with a LC as the cavity spacer layer and with CsPbBr$_3$ microplates inserted as the gain material \cite{21-xiong condensation, Zhu perovskite}.  If an electric field is applied to the microcavity, the polariton modes of the system can be tuned by virtue of the rotation of the LC molecular director as indicated in Figure 1(a). Increasing the pumping intensity, polariton condensation occurs in the form of a vortex mode with the topological charge of +1 under the voltage of 1 V. With strengthening the electric voltage to 3.6 V, the polariton condensate transforms into a vortex with the topological charge of +2. The topological charge of the vortex can be even transformed to -2 and -1 with further increase of the voltage to 7 V and 10 V, respectively, where the LC molecule is rotated nearly along the growth direction of the microcavity and the orthorhombic structure of the CsPbBr$_3$ microplate flips the optical anisotropy in the microcavity plane direction \cite{CsPbBr anisotropy, BIT perovskite}, cf. Figure 1(b). Our results show how the topological charge of the polariton vortex can be controlled by a simple electrical method.  The vortex control we demonstrate in the polariton many-body condensate is not only of fundamental interest but also paves the way to the realization of room-temperature operated micron-sized vortex elements acting as light emitters with controllable orbital angular momentum for applications in information processing.  
	
	
	\begin{figure}[ht]
		\centering
		\includegraphics[width=\linewidth]{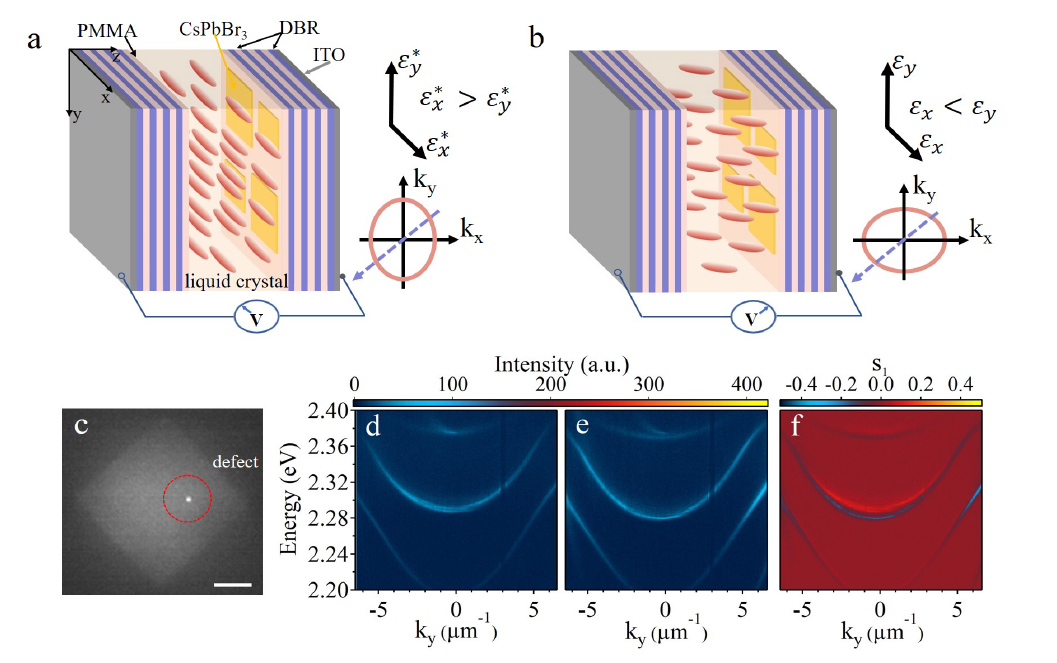}
		\caption{\textbf{Schematic diagram of the LC microcavity and properties of the perovskite microcavity without LC.} (a) The LC molecules are aligned along \textit{x} direction at low applied voltage, with permittivity $\varepsilon_x^{*}>\varepsilon_y^{*}$
			. In the anisotropic environment with $m_x < m_y$, the polariton mode distribution in momentum space ($k_x$, $k_y$ along \textit{x} and \textit{y} direction in (a)) is indicated in the lower right corner of the panel. The dashed arrow represents the potential gradient. (b) At larger voltage the LC molecular director is rotated to be nearly vertical to the microcavity plane. The anisotropy of the perovskite microplates results in the opposite permittivity distribution with $\varepsilon_x<\varepsilon_y$ and polariton mode distribution as indicated with $m_x > m_y$. (c) Near field imaging of the perovskite microplate below threshold with defect as indicated. The scale bar is 10 $\mu$m. (d) Horizontally and (e) vertically linearly polarized dispersion of the perovskite microcavity without LC (line profile  is plotted in the SM). (f) Calculated Stokes parameter $S_1$ of the dispersion.}
	\end{figure}
	
	
	
	The microcavity is fabricated using similar procedure as \cite{Yao RD}. We select one CsPbBr$_3$ microplate where a small defect is observed (Figure 1(c)) that occurs unintentionally during the growth process and it can be regarded as a potential trap (Figure S1). In addition, the two DBRs are tilted such that the microcavity length is varied slightly between different spatial positions and the potential gradient is along the direction of 60$^{\circ}$ against the \textit{x} axis (Figure S2). The appearance of the small defect and the potential gradient in the microcavity benefits the formation of vortices when the pump is above threshold. 
	
	We observe a clear XY splitting before the LC molecule is filled into the microcavity (Figure 1(d-f)), confirming the existence of the anisotropy of the perovskite microplate. The higher energy of the horizontally linearly polarized polariton modes shows that the permittivity of the perovskite microplate along \textit{x} direction is smaller than that along \textit{y} direction, i.e., $\varepsilon_x<\varepsilon_y$. We rub the ordering layer along \textit{x} direction and use positive type LC E7 ($\varepsilon_x^{*}$ is larger than $\varepsilon_y^{*}$ when the molecule is aligned along \textit{x} direction under zero voltage) such that the permittivity distribution of the LC molecule is opposite to the perovskite microplate. Under a small voltage, the anisotropy from the LC molecule dominates, while it takes a negligible role when the voltage is large enough, i.e., when the LC molecule is rotated to be nearly along the growth direction of the microcavity. Under larger voltages, the anisoptropy of the perovskite microplates plays an important role in the polariton condensation process.


	We measure the dispersion of the LC microcavity below threshold at the voltage 1 V, plotted in Figure 2(a) using a femtosecond laser with the repetition rate of 5700 Hz. Multiple lower polariton branches LP-1-LP3 are observed due to the large cavity length, that is, several cavity modes exist within the microcavity and strongly couple with the excitons in the perovskites (Figure S3). The strong coupling is confirmed at larger wavevectors where one can see clearly the dispersion bending. The polariton dispersions are fitted by using a coupled oscillator model and the details are shown in Table S1. The electric field rotates the LC molecule director, so that the horizontally linearly polarized modes LP-1, LP1 and LP3 in Figure 2(a) can be tuned to higher energy, while the vertically linearly polarized modes LP0 and LP2 remain fixed  \cite{1-liquid crystal_science, Yao RD}. 
	
	
	As increasing the pumping intensity to around 12 $\mu$J/cm$^2$ under the voltage of 1 V (well below the Mott density 40 $\mu$J/cm$^2$\cite{Zhu perovskite}), the emitted intensity from LP1 branch, which is well below the cavity modes, increases superlinearly. Whereas its linewidth drops suddenly from 6 meV to 1 meV, as shown in Figure 2(b) and Figure S5(a-c), confirming the occurrence of polariton condensation (the energy blueshift can be observed below and above the threshold, see Figure S5(d)). Due to the nonuniformity of the perovskites, the polaritons condense locally in an area with the size of around 8 $\mu$m. As expected, an intensity minimum in the center of the polariton condensate can be clearly seen (Figure 2(c)) due to the existence of the small defect shown in Figure 1(c).
	
	\begin{figure}[t]
		\centering
		\includegraphics[width=\linewidth]{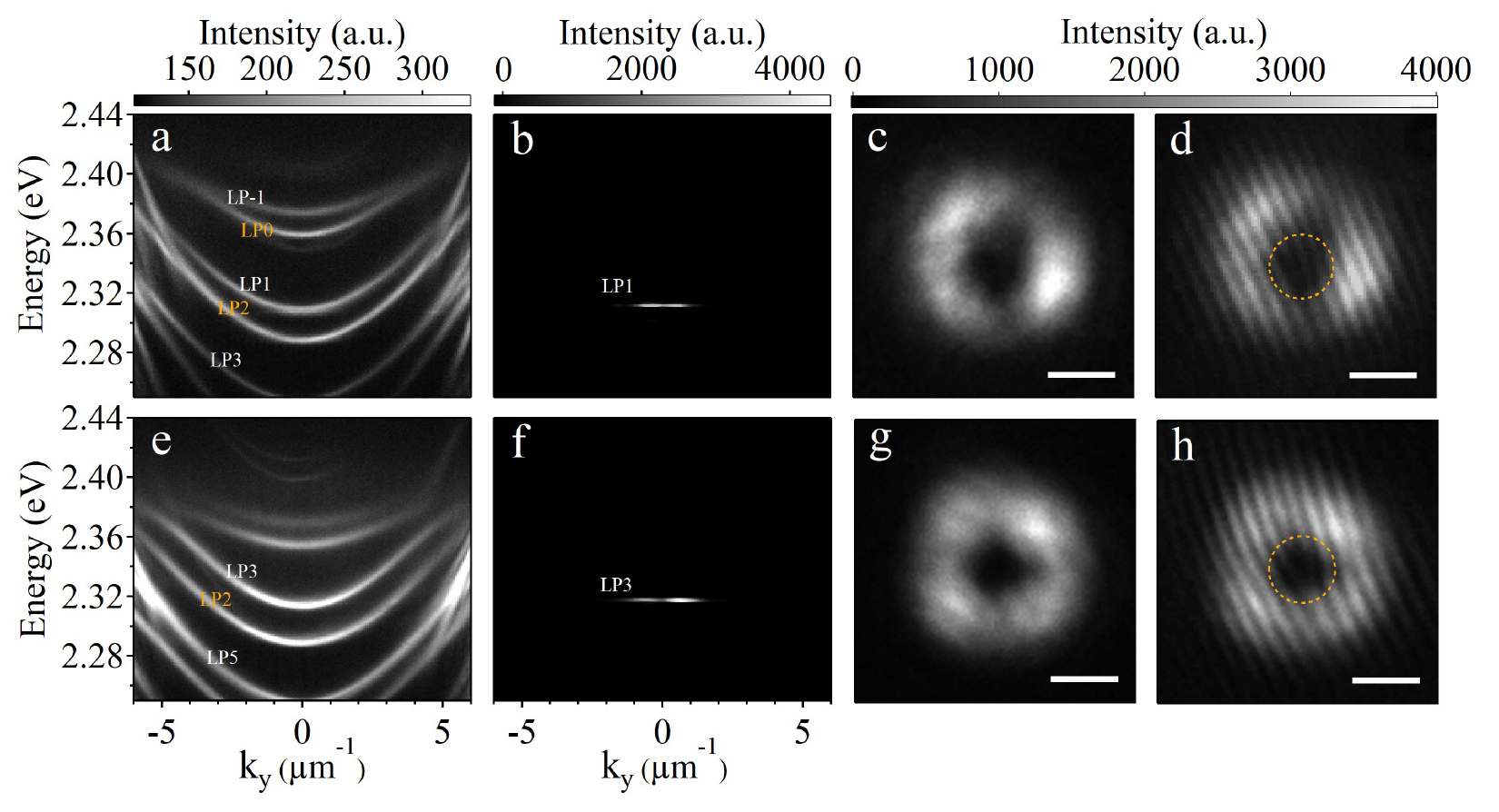}
		\caption{\textbf{Vortices at smaller voltages.} {(a, b) Dispersion below and above threshold under the voltage of 1 V. (c) Real space imaging of the polariton condensate corresponding to (b). (d) Interferogram of the polariton condensate in (c), indicating that the topological charge is +1. (e, f) Dispersion below and above threshold under the voltage of 3.6 V. (g) Real space imaging of the polariton vortex corresponding to (f). (h) Interferogram of the polariton condensate in (g), indicating that the topological charge is +2. The dashed lines in (a, b, e, f) represent the cavity modes of the corresponding polariton branches spectrally well below them. The inserts in (d, h) are the corresponding phase information, obtained by Fourier transform of the interferograms of the area indicated by the dashed squares.} The scale bars: 3 $\mu$m. 
		}
	\end{figure}
	

	To demonstrate the existence of the macroscopic coherence and figure out the phase distribution of the ring-shaped polariton condensate, we build a Michelson interferometer with one arm being expanded by around 15 times. By superimposing the real space image of the polariton condensate onto the expanded area where the phase distribution is nearly uniform, one can clearly see the macroscopic coherence above threshold (Figure 2(d)). A fork with two fringes can be clearly observed in the intensity minimum region, confirming the existence of a vortex core with the topological charge of +1. The presence of the vortex is further supported by the spectrum shown in Figure 2(b). The intensity peak is located at nonzero momenta of $\pm$ 0.67 $\mu$m$^{-1}$ caused by the finite OAM, consistent with the fork appearing in the interferogram graph. {Note that the small voltage applied here does not influence the distribution or dynamics of the condensate, which means that the vortex can still be excited in the absence of the electric field.}
	
	\begin{figure}[t]
		\centering
		\includegraphics[width=\linewidth]{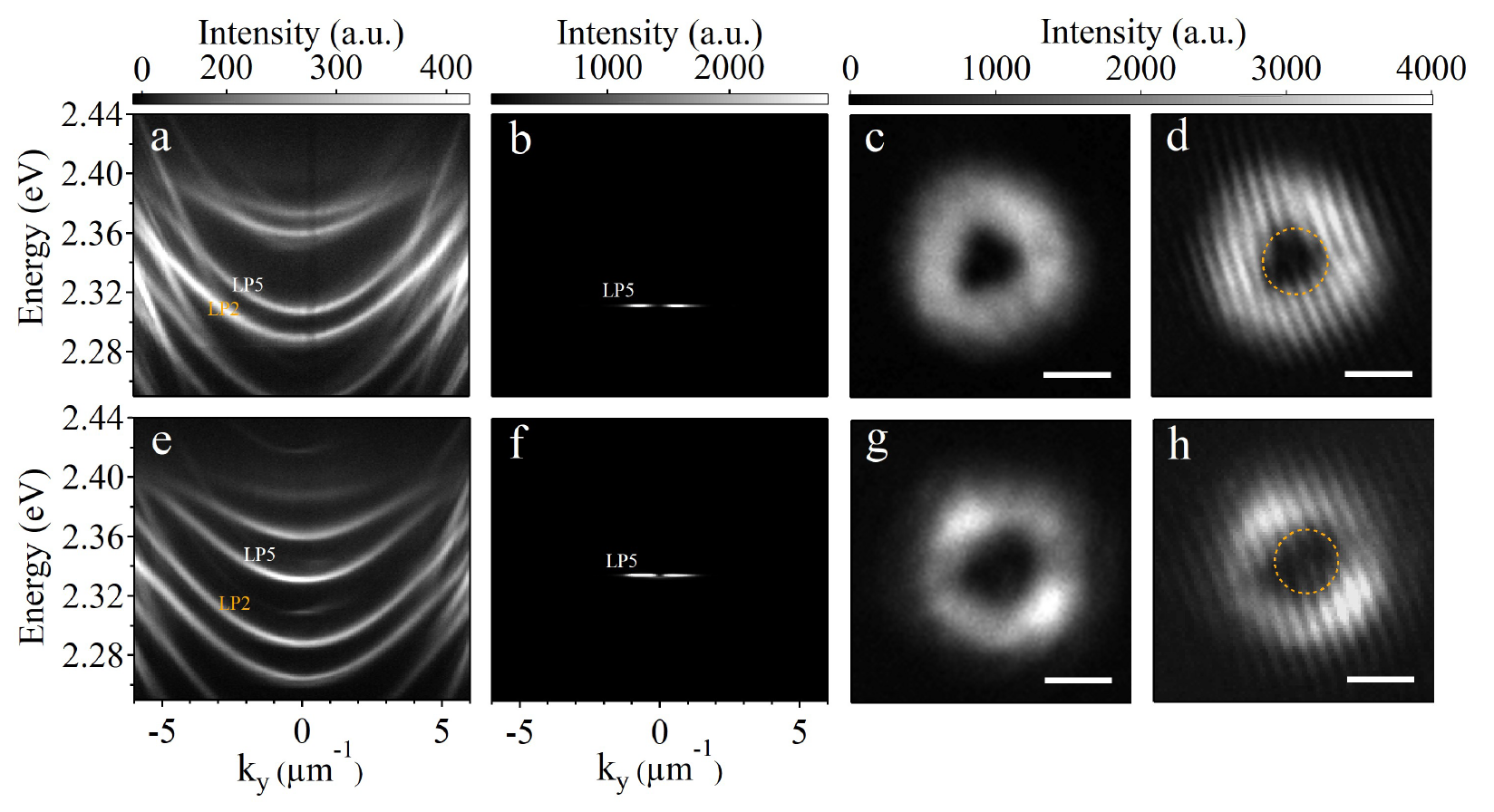}
		\caption{\textbf{Vortices at larger voltages.} (a, b) Dispersion below and above threshold under the voltage of 7 V. (c) Real space imaging of the polariton vortex corresponding to (b). (d) Interferogram of the polariton condensate in (c), indicating that the topological charge is -2. (e, f) Dispersion below and above threshold under the voltage of 10 V. The signal between LP2 and LP5 is due to the PL out of the perovskites. (g) Real space imaging of the polariton vortex corresponding to (f). (h) Interferogram of the polariton condensate in (g), indicating that the topological charge is -1.The dashed lines in (a, b, e, f) represent the cavity modes of the corresponding polariton branches spectrally well below them. The inserts in (d, h) are the corresponding phase information of the area indicated by the dashed squares. The scale bars: 3 $\mu$m. }
		
	\end{figure}
	
	Increasing the voltage further can bring the LP1 branch to the higher energy with the polariton emission intensity being greatly reduced. On the other hand, the LP3 branch is blueshifted to beyond the LP2 branch (Figure 2(e, f)). During this process, a four-lobe state appears (Figure S6) in a very small voltage range. Similar multi-lobe modes in polariton condensates have also been observed in resulting from the asymmetry of the pump profile\cite{WGM CdTe}. For the anisotropy in the current work, the four-lobe mode disappears quickly and transforms further to a new vortex mode under the voltage of 3.6 V (Figure 2(g)). After examining the polariton interferogram in Figure 2(h), we observe a fork in the density minimum area of the polariton condensate with three fringes, manifesting that the topological charge of the vortex is +2. Generally, the higher-order vortices have a larger size, in both momentum and real spaces, compared with the fundamental ones. From the spectrum taken under the voltage of 3.6 V (Figure 2(f)), one can see that the peak of the condensate in momentum space is increased to $\pm$ 0.85 $\mu$m$^{-1}$. This is larger than the peak momentum under the voltage of 1 V where a vortex of charge +1 forms and simultaneously demonstrates that a higher-order topological charge is excited. The reason why the higher-order vortex is excited at this voltage is that in polariton systems, the vortices are formed spontaneously under non-resonant excitation and closely affected by the effective mass, detuning, and lifetime of polaritons. As the condensate jumps from LP1 to LP3 (Figure 2(b,f)) which has a larger effective mass (Table S2 in the SM and \cite{parameters1} ), the polaritons condense spontaneously to the higher-order vortex state (also see the numerical results below).
	
	\begin{figure}[t]
		\centering
		\includegraphics[width=\linewidth]{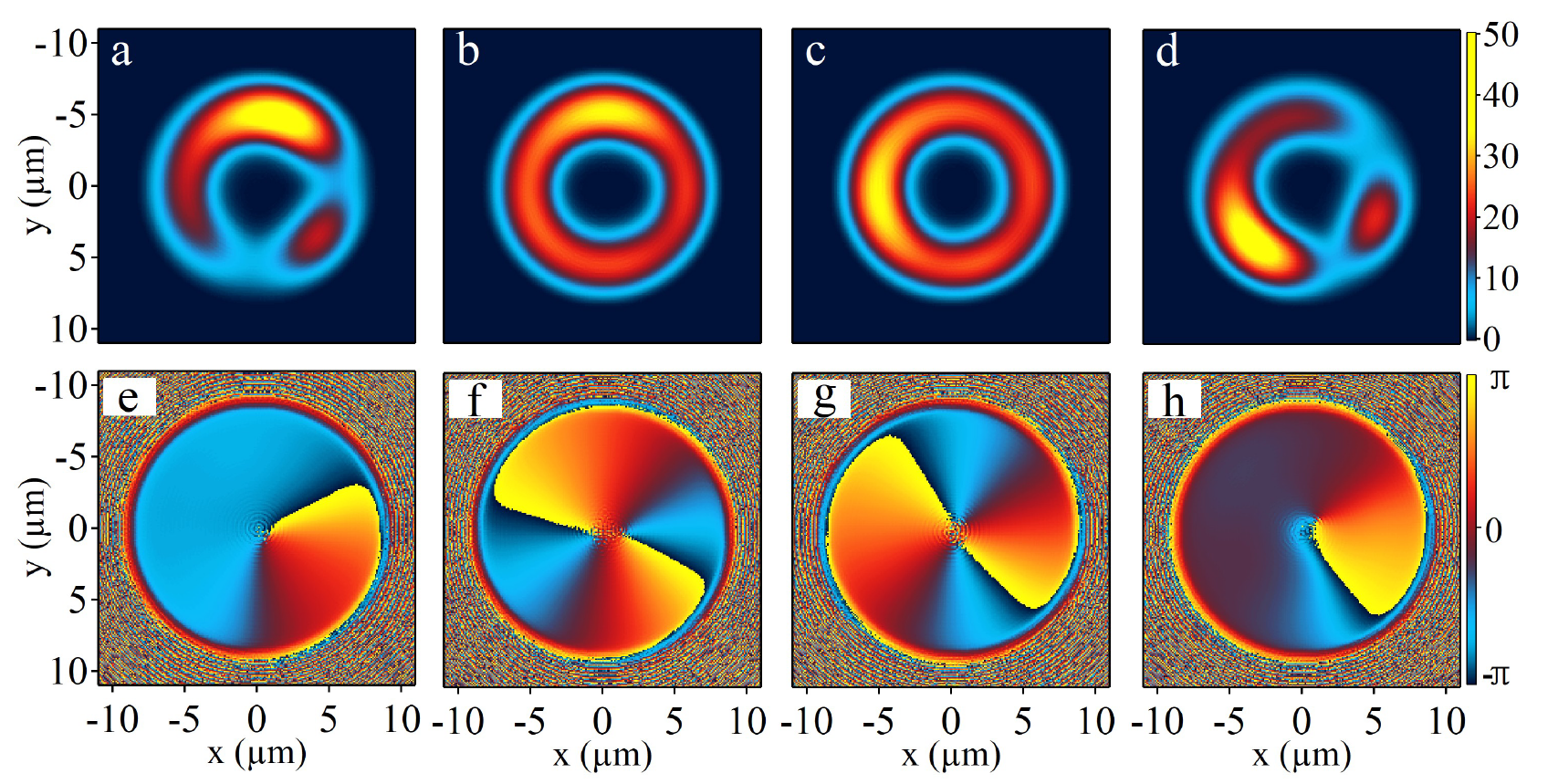}
		\caption{{\textbf{Numerical results of stable vortices.} Density (upper row, in $\mu$m$^{-2}$) and phase (lower row) profiles of stationary vortices obtained in numerical simulations for a fixed point in time. Results in (a-d) correspond to the experimental cases obtained at voltages of 1~V, 3.6~V, 7~V, and 10~V, respectively. The parameters are: $\gamma_\textup{c}=0.4$ ps$^{-1}$, $\gamma_\textup{r}=$1.5 $\gamma_\textup{c}$, $g_\textup{r}=$2 $g_\textup{c}$, and $R=0.01$ ps$^{-1}$. The effective masses \cite{parameters1} are extracted from the experiments. According to the reported exciton-exciton interaction strength~\cite{exciton-interaction1, exciton-interaction2, exciton-interaction3}, the polariton-polariton interaction strength $g_\textup{c}$ has been estimated as in~\cite{parameters2}}.}
	\end{figure}
	
	It is worth noting that higher-order vortices in polariton condensates in planar microcavities are unstable due to the strong polariton-polariton interaction. A vortex with topological charge +2 will split into two singly charged vortices with two cores or simply jump to the lower-order vortex state with topological charge +1, depending on the pumping condition \cite{Sanvetto vortex 2, resonant vortex 2}. In our experiments, the polariton condensate is mainly located within an area of around 8 $\mu$m on one corner of the CsPbBr$_3$ microplate, so that the higher-order vortex can be stabilized by the finite size of the system. We note that the difference of the size of the vortex cores in Figure 2(c,g) is very small, because they originate from the existence of the small defect and the finite size of the microplate. Similar same-sized vortices with different topological charges are reported in a ring-shaped potential trap \cite{Ma ring shape}. 
	
	Under higher voltages, the LC molecule can be rotated with a large angle to be nearly perpendicular to the planar surface. As a result, the LC-induced anisotropy in the microcavity can be neglected, and the anisotropy (Figure 1(d, e, f)) in the microcavity caused by the orthorhombic symmetry of the CsPbBr$_3$ microplate dominates and influences the polariton modes \cite{CsPbBr anisotropy}. At voltage of 7~V, the near field of the polariton condensate shows a density minimum and a three-fringe fork (Figure 3(d)), which is towards the opposite direction of that shown in Figure 2(h). Therefore, it is clear that a vortex with the topological charge of -2 is formed, as shown in Figure 3(a-d). A vortex with the topological charge of -1 is observed when the voltage is further increased to 10~V, as shown in Figure 3(e-h). The transition between the two vortices is smooth and no intermediate modes are observed. The peak momentum of the polariton condensate decreases from $\pm$ 0.79 $\mu$m$^{-1}$ to $\pm$ 0.73 $\mu$m$^{-1}$ when the voltage is increased from 7~V to 10~V, opposite to the vortex transition from 1~V to 3.6~V, which is consistent with the variation of the topological charge.

	The manipulation of the topological charge from +2 to -2 originates from the tunability of the anisotropy of the microcavity. We calculate the effective mass $m_x$ along $k_x$ direction and $m_{y}$ along $k_y$ direction at the bottom of the polariton branches LP1, LP3 and LP5 at different voltages. {We find that $m_{x}$ of the polariton branch LP1 and LP3 is smaller than $m_{y}$ at the voltage of 1~V and 3.6~V, whereas at the bottom of LP5, $m_{x}$ is larger than $m_{y}$ at the voltage of 7~V and 10~V (see \cite{parameters1} and SM.)  In other words, the reversed ratio of $m_{x}$ and $m_{y}$ gives rise to the modulation of the topological charge of the vortex. The vortex with the topological charge of -2 becoming unstable when the voltage is increased to 10~V attributes to the increased interaction strength between exciton reservoir and condensate.
		
		To provide further insight into the measured results shown above, we numerically simulate the dynamics of polariton condensates based on the Gross-Pitaevskii equation: 
		\begin{equation}
		\begin{aligned}\label{eq:GP}
		i\hbar\frac{\partial\Psi(\mathbf{r},t)}{\partial t}=&\left[-\frac{\hbar^2 \Delta_{x}}{2m_{x}}-\frac{\hbar^2 \Delta_{y}}{2m_{y}}-i\hbar\frac{\gamma_\text{c}}{2}+g_\text{c}|\Psi(\mathbf{r},t)|^2 \right. \\
		&\left.+\left(g_\text{r}+i\hbar\frac{R}{2}\right)n(\mathbf{r},t)+V(\mathbf{r})\right]\Psi(\mathbf{r},t)\,.
		\end{aligned}
		\end{equation}
		Here, $\Psi(\mathbf{r},t)$ is the polariton field and $\Delta$ is the Laplacian operator. $m_{x(y)}$ describes the effective mass of polariton condensates along $x(y)$ direction and they can be electrically controlled by tuning the orientation direction of the LC molecules. In quasi-mode approximation~\cite{AppliedOptics} the non-equilibrium polariton condensate decays with a rate $\gamma_\text{c}$ and is replenished by reservoir excitons with a condensation rate $R$. $g_\text{c}$ represents the polariton-polariton interaction and $g_\text{r}$ represents the polariton-reservoir interaction. The finite size of the perovskite microplate and the small defect give rise to a trapping potential $V(\mathbf{r})$ (see SM for the details). Simultaneously, a small potential gradient along an angle between $x$ and $y$ axes is added to break the rotational symmetry of the system in the case of $m_x\neq{m_y}$, to create a specifically charged vortex. The density of the reservoir obeys the following equation of motion:
		\begin{equation}\label{eq:reservoir}
		\frac{\partial n(\mathbf{r},t)}{\partial t}=\left[-\gamma_r-R|\Psi(\mathbf{r},t)|^2\right]n(\mathbf{r},t)+P(\mathbf{r})\,.
		\end{equation}
		Here, $\gamma_\text{r}$ is the loss rate of the reservoir and $P(\mathbf{r})$ is the non-resonant pump with a super-Gaussian shape and a broad size covering the entire potential (the details are given in the SM). Note that the pump in the numerical simulation is a continuous wave. This lets us further demonstrate that the observed vortices are stationary solutions of the nonlinear system.   
		
		From the numerical result shown in Figure 4(a, e), one can see that a vortex with topological charge +1 is excited, because the exciton reservoir in the defect hinders the condensation of the polaritons, leaving a density minimum, and the specific topological charge is due to the appearance of the potential gradient and the anisotropic effective mass ($m_x<m_y$). As the effective mass increases with the electric field, the condensate builds up spontaneously to a vortex state with a higher topological charge +2 (Figure 4(b, f)). The positively charged vortex can be tuned to rotate towards the opposite direction carrying a negative charge when the system's asymmetry is spatially mirrored. This can be achieved by changing the ratio of the effective mass along $x$ and $y$ directions. When $m_x>m_y$, as shown in Figure 4(c, g), the vortex is flipped to carry a topological charge -2. Further increasing the electric field enhances the polariton-reservoir interaction (see \cite{parameters2} and SM) as well as the condensation rate, leading to the decay of the higher order topological charge and consequently the formation of a vortex with charge -1 (Figure 4(d, h)). 
		
		To summarize, we observe polariton vortex formation in a LC microcavity with CsPbBr$_3$ microplates acting as the gain material. Thanks to the tunable anisotropy of the refractive index of the microcavity, the topological charge of the vortex can be tuned from +1 to +2, and then can be manipulated to -2 and further to -1 by simply increasing the voltage applied to the microcavity with all other system and excitation parameters remaining unchanged. Our results may find important application in vortex-based information processing in microscale systems operating at room temperature. The control process of the quantized vorticity realized in our system can be easily integrated into optoelectronic and photonic circuits.
		
		\begin{acknowledgments}
			TG acknowledges the support from the National Natural Science Foundation of China (grant Nos. $12174285$). XM acknowledges funding from the Deutsche Forschungsgemeinschaft (DFG) (grant No. 467358803). The Paderborn group acknowledges a grant for computing time at Paderborn Center for Parallel Computing, PC$^2$.
		\end{acknowledgments}


\begin{thebibliography}{99}
\bibitem{parameters1} The measured effective masses $m_x$ and $m_y$ at 1~V, 3.6~V, 7~V and 10~V are  \newline
			$m_x:1.19\times10^{-5} m_e,\ m_y:1.41 \times10^{-5}m_e$; \newline 
			$m_x:1.24\times10^{-5} m_e,\ m_y:1.51 \times10^{-5}m_e$; \newline
			$m_x:1.23\times10^{-5} m_e,\ m_y:1.16 \times10^{-5}m_e$; \newline
			$m_x:1.33\times10^{-5} m_e,\ m_y:1.26 \times10^{-5}m_e$, respectively.
\bibitem{parameters2} The 2D exciton-exciton interaction strength reported in \cite{exciton-interaction1, exciton-interaction2} is around 4.1 $\mu$eV $\mu$m$^2$ in CsPbBr$_3$, so that the interaction strength in 3D in our sample (219 nm thick) is around 0.9 $\mu$eV $\mu$m$^3$. The polariton-polariton interaction strength $g_\textup{c}$ can be calculated by considering the amount of exciton component, which leads to 0.0044 $\mu$eV $\mu$m$^3$ (1~V), 0.0078 $\mu$eV $\mu$m$^3$ (3.6~V), 0.00182 $\mu$eV $\mu$m$^3$ (7~V), and 0.01496 $\mu$eV $\mu$m$^3$ (10~V).
\end{thebibliography}
	\end{document}